\documentclass[12pt]{article}

\usepackage{graphics}
\usepackage[dvips]{graphicx}

\begin{document}

{\footnotesize jcis@epacis.org}

\begin{center}

{\bf Programing Using High Level Design With Python and FORTRAN: A Study Case in Astrophysics}
\bigskip


{\small Eduardo S. Pereira\footnote{duducosmo@das.inpe.br}, Oswaldo D. Miranda \footnote{oswaldo@das.inpe.br}
}
\smallskip

{\small
Instituto Nacional de Pesquisas Espaciais - Divis\~{a}o de Astrof\'{i}sica,\\
Av. dos Astronautas 1758, S\~{a}o Jos\'{e} dos Campos, 12227-010 SP, Brazil
}

{\footnotesize Received on September , 2011 / accepted on *****, 2010}

\end{center}


\begin{abstract}

In this work, we present a short review about the high level design methodology (HLDM), that is based on the use of very
high level (VHL) programing language as  main, and the use of the intermediate level (IL) language only for the critical
processing time. The languages used are Python (VHL) and  FORTRAN (IL). Moreover, this methodology, making use of the oriented
object programing (OOP), permits to produce a readable, portable and reusable code. Also is presented the concept of computational
framework, that naturally appears from the OOP paradigm. As an example, we present the framework called PYGRAWC (Python framework for
Gravitational Waves from Cosmological origin). Even more, we show that the use of HLDM with Python and FORTRAN produces a powerful tool for solving
astrophysical problems.

\bigskip

\end{abstract}

\textbf{1. INTRODUCTION}
\bigskip
\bigskip

Python \cite{Pythonof,pybook} is a very-high level dynamic language programming. The Python interpreter is available for different operational systems (OS).
This means that is possible to write a code which can run in different OS without requiring any modification. However, in order to
have the best performance, the critical computational part of the code should be written in a compiled language like c/c++ or FORTRAN. This way of writing
codes is called of high level design (HLD) \cite{hinsenetal,kh}. Another important fact is that, in general, $80$\% of the runtime is spent in
$20$\% of the code (Pareto Principle) \cite{Behenel11}. Thus, the use of a VHL for the principal part of the code permits a more agile processing.
Although the HLD goes beyond of a mixing of different languages, an important feature of HLD is related to the use of oriented object programming paradigm,
OOP, working together an Unified Modeling Language (UML) class diagram.

\bigskip
\bigskip

\textbf{2. The High Level Design}
\bigskip
\bigskip

The first step to write an efficient code consists in dividing the problem in classes. In this section we present programs that make this task easier,
as for example, \textit{dia} \cite{dia} and \textit{dia2code} \cite{d2c}. The first one permits us to write class diagrams,  and with the second program
we can generate a frame code. That is, the class in the diagram image which is converted to a class in code structure. Through this
section these concepts will become clearer.

\bigskip
\textbf{2.1 Planning Before Programing }
\bigskip

We start this section showing an example of class diagram\footnote{All documentation about how to install and use the {\it dia} software can
be found in \cite{dia}.}. Through this paper, we will use an example derived from {\it cosmology}. In particular, the main characteristics
of a cosmological model are: the age of the Universe, the scale factor which describes how the radius of the Universe evolves with time, and
the density of matter/energy.

\begin{figure}[h!]
	\begin{center}
		{\resizebox{0.5\columnwidth}{!}{\includegraphics{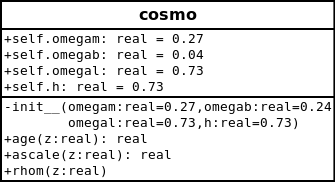}}}
	\end{center}
	\caption{\textit{cosmo.dia}, an example of class diagram for the basic characteristics of a cosmological model.}
	\label{fig:2}
\end{figure}

The class diagram that represents this cosmological model is showed in the figure \ref{fig:2}. The attributes of this class
are the cosmological parameters, at the present, for total matter (self.omegam - $\Omega_{m}$), barionic matter (self.omegab -$\Omega_{b}$),
dark energy (self.omegal - $\Omega_{\Lambda}$) and Hubble parameter (self.h- $h$ )\footnote{The Hubble constant at the present time
is written in terms of $h$ by $H_{0}=100\,h\,{\rm km}\,{\rm s}^{-1}\,{\rm Mpc}^{-1}$ (where $1\,{\rm Mpc}=3.086\times 10^{24}\, {\rm cm}$).}.

The file is saved with the name \textit{cosmo.dia}. Now, it is possible to generate a structured code with \textit{dia2code} (see \cite{d2c} for details),
using the command \textit{dia2code cosmo.dia -t Python}. The code generated by this example, and all examples used in this article, can be downloaded
from \cite{dudu2}. It is possible to generate c++ and java structured code choosing the name of the equivalent language from the \textit{dia2code}
command.


However, this way to structure a code class is only a start point. It is necessary to do a better organization and fill the methods with
the equivalent operations.

\bigskip
\bigskip

\newpage
\textbf{2.2 Mixing Python and FORTRAN}

\bigskip
\bigskip

There is a very useful tool, called \textit{f2py} \cite{pearu09}, which permits to do a wrapper of a FORTRAN 77 code to Python. That is,
it compiles the FORTRAN subroutine in a format which can be used by Python module. The \textit{f2py} is contained in the package \textit{numpy}
\cite{scipy}. Below, we present a simple example:

\bigskip
C FILE hiword.f

~~~~~~subroutine hiword(a,b)
       
~~~~~~~~~~real*8 a,b
       
cf2py intent(in) a

cf2py intent(out) b

~~~~~~~~~~b = a*a
       
~~~~~~~~~~write(6,*) 'b = ',b,', a = ',a
       
~~~~~~~~~~return
       
~~~~~~end
       
\bigskip
The comment \textit{cf2py} allows the \textit{f2py} wrapper can be identified with both the input and output variables in the function \textit{hiword}.
Giving the name \textit{hiword.f} to the file contained in the above code, we can compile it from the following command:
\bigskip

f2py -c hiword.f -m hiword

\bigskip
In this case, the \textit{-c} means compile, and \textit{-m} generate a Python module with name \textit{hiword}. Below, we present an example
how to call the function \textit{hiword} in a Python code.

\bigskip
$[1]>>>$import hiword

$[2]>>>$print hiword.\_\_doc\_\_

$[3]>>>$ This module 'hiword' is auto-generated with f2py (version:2).

$[4]>>>$Functions:

$[5]>>>$ b = hiword(a)

$[6]>>>$hiword.hiword(5)

$[7]>>>$ b =    25.000000000000000      , a =    5.0000000000000000     

$[8]>>>$ 25.0

\bigskip
The text in front of $>>$ represents what is printed in the display. For more details and examples see \cite{pearu09}.

\bigskip
\bigskip

\textbf{2.2 Optimizing the Code for Multi-Core Machines}

\bigskip
\bigskip

Another interesting fact about Python is that it has a lot of modules. One of this is the \textit{multiprocessing} that permits to write a parallel
code in an easy way. As an example of using this module in scientific computing, consider the following equation:

\begin{equation}\label{eq:1}
f(x) = \int_{a}^{b}{g(x,k)}dk,
\end{equation}

\noindent where $a \leq k \leq b$.

\bigskip

\begin{figure}[h!]
	\begin{center}
		{\resizebox{0.5\columnwidth}{!}{\includegraphics{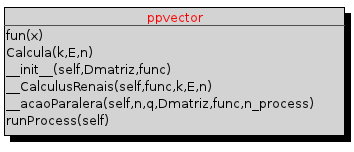}}}
	\end{center}
	\caption{\textit{ppvector}, a Python module for construction of parallel scientific code in a multi-core machine.}
	\label{fig:3}
\end{figure}

In many cases $g(x,k)$ can not be written in a separated form. In this case, the integral equation must be evaluated for each $x$
in a given range $[x_{0},x_{f}]$. However, we can divide the range $[x_{0},x_{f}]$ by the number of central processor units (CPU)
of a cluster compute (or multi-core machine), and so we can calculate $f(x)$ in parallel mode. In the figure \ref{fig:3}, it is showed
the class diagram of \textit{ppvector}, that is a class we developed to do this type of operation in parallel model, for multi-core
machine, based on the module \textit{multiprocessing}. The source code can be downloaded from \cite{dudu}.

\bigskip

The code below shows the use of\textit{ ppvector}:


\bigskip
import multiprocessing as mpg

from ppvector import ppvector

from scipy.integrate import romberg

np=10000;   zmax=20.0;    deltaz=zmax/np

g= mpg.Array('d',[0 for i in range(np)]) \# The d indicate duble precision

z= mpg.Array('d',[zmax-i*deltaz for i in range(np)])

\#Define a function that will be  calculate the integral in parallel

\#k  is the starter point of the sub-range

\#E  is the lenght of the range

\#n  is the number of CPU's of machine

def f(x):

~~~~def f2(k):

~~~~~~~~return (x+k)**(-2.0)
        
~~~~return romberg(f2,5.0,20.0)

def fun(k,E,n):

~~~~k2=k+E
    
~~~~for i in range(k,k2+1):
    
~~~~~~~~zloc=z[k]
        
~~~~~~~~g[k]=f(zloc)

C1= ppvector(np,fun) \# Star the ppvector class

C1.runProcess() \# Executing the parallel calculus.

\bigskip

\bigskip

The function \textit{Array}, of \textit{multiprocessing} module, allocates a matrix in a global memory which can be accessed by all CPU's.
In line $25$ is passed on the length of the vector and the function that divides the job in sub-ranges. In line $26$, the parallel code
is called and executed.

\bigskip
\bigskip

\textbf{3. Python Framework for Cosmological Gravitational Waves - PYGRAWC}

\bigskip
\bigskip

A framework is a set of classes, interfaces and patterns to solve a group of problems. It is like a little application with statical and
dynamical structures to solve a set of restrict problems. So, a framework is more than a simple library (we refer the reader to
\cite{fayad00,fayad97,govoni99}).

\bigskip

\begin{figure}[h!]
	\begin{center}
		{\resizebox{0.8\columnwidth}{!}{\includegraphics{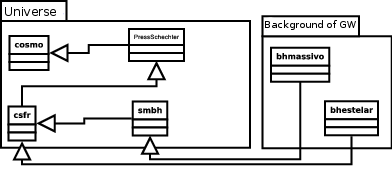}}}
	\end{center}
	\caption{The \textit{PYGRAWC} framework class diagram.}
	\label{fig:4}
\end{figure}

In figure \ref{fig:4} is presented the class diagram of the core of \textit{PyGraWC}. It is a framework that we are developing to study
gravitational waves from cosmological origin. Here, it is only showed the class name and the relation among their several components.

\bigskip

The class \textit{cosmo} describes the background cosmology. The class \textit{PressSchechter} is based on a Press-Schechter-like formalism
\cite{pereira2011} and it describes both the evolution of dark matter halos and the infall of barionic matter in these halos.
The class \textit{csfr} describes the evolution of the cosmic star formation rate. The class \textit{smbh} describes the evolution of
supermassive black holes in the centers of galaxies. The classes \textit{bhestelar} and \textit{bhmassivo} calculate the stochastic background
of gravitational waves generated by: the collapse of stars to form black holes \cite{pereira2011} and the growth of supermassive black holes
(in progress). All details about the astrophysical model and the results obtained from this framework can be seen in
\cite{pereira2011,pereira,PereiraB:2010}.

\bigskip
\bigskip
\newpage
\textbf{4. Final Considerations}

\bigskip
\bigskip

In this work is presented a High Level Design methodology (HLD), that consists in the mixing of a very-hight level interpreted language (VHL)
with an intermediated compiled language (IL). Using tools of software engineering, like UML, and also framework concept, we can write efficient
scientific codes without spending a lot of time in the development phase. Here, it was used Python (VHL) and Fortran (IL) and it was showed
that this combination can be easily done giving excellent results, as can be seen by the presentation of Python framework for Gravitational
Waves from Cosmological origin (\textit{PyGraWC}).

\bigskip
\bigskip
{Acknowledgments: E.S. Pereira would like to thank the Brazilian Agency CAPES for support. O.D. Miranda would like to thank the
Brazilian Agency CNPq for partial support (grant 300713/2009-6)}

\end{document}